\documentstyle[epsf,astrcite]{mn}
\newcommand{\un}[2]{\mbox{\rm\thinspace #1$^{#2}$}}

\newcommand{\be}[1]{\begin{equation}\label{#1}}
\newcommand{\ee}{\end{equation}}

\newcommand{\degs}{\mbox{$^{\circ}$}}
\newcommand{\Eq}[1]{Eq.\,(\ref{#1})}

\newcommand{\Fig}[1]{Fig.\,\ref{#1}}

\newcommand{\gsim}{\mathrel{\hbox{\rlap{\lower.55ex \hbox {$\sim$}}
                   \kern-.3em \raise.4ex \hbox{$>$}}}}
\newcommand{\lsim}{\mathrel{\hbox{\rlap{\lower.55ex \hbox {$\sim$}}
                   \kern-.3em \raise.4ex \hbox{$<$}}}}
\newcommand{\msun}{\mbox{M$_\odot$}}

\newcommand{\sect}[1]{sect.\,\ref{#1}}

\newcommand{\sub}[1]{_{\rm #1}}

\newlength{\ffh}

\def\farcs{\hbox{$.\!\!^{\prime\prime}$}}

\newcommand{\dhat}{\mbox{$\hat{d}$}}
\title[Fallback bubble around Vela pulsar]
    {The Vela pulsar `jet':\\ a companion-punctured bubble of fallback material}
\author[R. Wijers and S. Sigurdsson]
       {Ralph A.M.J. Wijers and Steinn Sigurdsson\\
        Institute of Astronomy, Madingley Road, Cambridge CB3 0HA, UK\\
	E-mail: {\tt ramjw@ast.cam.ac.uk} and {\tt steinn@ast.cam.ac.uk}}
\date{\underline{submitted 17 December 1996, revised 17 April 1997}}

\begin{document}

\maketitle

\begin{abstract}
Markwardt and \"Ogelman (1995)\nocite{mo:95} used {\it ROSAT\/} to reveal
a 12 by 45 arcmin structure in 1\un{keV}{} X rays around the Vela pulsar,
which they interpret as a jet emanating from the pulsar. We here present
an alternative view of the nature of this feature, namely that it consists
of material from very deep inside the exploding star, close to the mass
cut between material that became part of the neutron star and ejected
material.  The initial radial velocity of the inner material  was lower
than the bulk of the ejecta, and formed a bubble of slow material that
started expanding again due to heating by the young pulsar's spindown
energy. The expansion is mainly in one direction, and to explain this we
speculate that the pre-supernova system was a binary. The explosion caused
the binary to unbind, and the pulsar's former companion carved  a
lower-density channel into the main ejecta.  The resulting puncture of the
bubble's edge greatly facilitated expansion along its path relative to
other directions. If this is the case, we can estimate the current speed
of the former binary companion and from this reconstruct the presupernova
binary orbit. It follows that the exploding star was a helium star, hence
that the supernova was of type Ib. Since the most likely binary companion
is another neutron star, the evolution of the Vela remnant and its 
surroundings has been rather more complicated than the simple expansion
of one supernova blast wave into unperturbed interstellar material.
\end{abstract}



   \section{Introduction}
   \label{intro}

      \subsection{Basic properties of Vela}
      \label{basic}

The Vela supernova remnant is a large radio supernova remnant (see Green
1988\nocite{green:88} and references therein)  associated with a nearly
circular X-ray shell with a radius of 4 degrees (Aschenbach et~al
1995)\nocite{aet:95}. It is centred on and physically associated with the
Vela pulsar (PSR\,B0833$-$45).  The traditional value of its distance is
500\un{pc}{} (Green 1988), but recent evidence both from optical
observations (Wallerstein et~al.\ 1995)\nocite{wvjf:95} and X-ray data 
(Aschenbach et~al.\ 1995; \Fig{fi:xray})\nocite{aet:95} are more consistent
with half that distance. Since various quantities relevant to our work do
depend on distance, we will quote their distance scaling where appropriate
in terms of
\begin{equation}
   \label{eq:dhat}
   \dhat\equiv d/250\un{pc}{},
\end{equation}
thereby implicitly adopting the closer distance as our standard value.

The pulsar has a spindown age $\tau\sub{sd}$ of 11\,000\un{yr}{}\ and a
spin period of 89\un{ms}{}, which implies a surface magnetic field of
3.4$\times10^{12}\un{G}{}$ and a total spin-down energy loss rate,
$L\sub{sd}$, of $7\times10^{36}\un{erg}{}\un{s}{-1}$ for canonical values
of neutron star parameters ($I=10^{45}\un{g}{}\un{cm}{2}$,
$R=10\un{km}{}$, $M=1.4\un{\msun}{}$). Most of the spindown energy loss is
not accounted for in any hitherto known emission or other sink of energy.
The pulsar is moving towards the northwest (position angle $-54\degs$)
with a proper motion of 59\un{mas}{}\un{yr}{-1} (Bailes
et~al.\ 1990)\nocite{bmknr:90}, which implies a transverse speed of
$70\dhat\un{km}{}\un{s}{-1}$. If we extrapolate back the pulsar position
to 11\,000\un{yr}{} ago, we find that it was then near a few dots of
enhanced emission near the east edge of the bubble, suggestive of this
being indeed the birth location of the pulsar.  Then the pulsar has a true
age close to the spindown age, hence its field has been approximately
constant over most of its past life and that the initial spin period was
significantly shorter than the current one.

      \subsection{The pulsar `jet'}
      \label{jet}

Markwardt and \"Ogelman (1995)\nocite{mo:95} found a 12 by 45 arcmin emission
region in the ROSAT 0.9--2.0\un{keV}{}\ band extending from the pulsar to
the south-southwest (\Fig{fi:xray}).  From bremsstrahlung fits to its spectrum, they
derived a temperature of $1.3\un{keV}{}/k$ and a density, $n\sub{b}$, of
$0.57\dhat^{-1/2}\un{cm}{-3}$. Frail et~al.\ (1996)\nocite{fbmo:96}
report that both ASCA data at higher energies and lower-energy ROSAT data
support the thermal nature of the emission.
\begin{figure*}
\begin{minipage}[b]{0.75\textwidth}
   {\Large\it see file }{\Large\tt vela\_fg1.gif}

   \vspace*{1cm}
\end{minipage}\hfill\begin{minipage}[b]{0.23\textwidth}
  \caption{The Vela supernova remnant in X rays (Aschenbach et~al.\ 
           1995) with the jet-like feature found by
           Markwardt \& \"Ogelman (1995) superposed on
           the same scale and enlarged in inset.
           \label{fi:xray}
           }
\end{minipage}
\end{figure*}

They interpreted this
feature as a hydrodynamic jet directly pushed by the pulsar, based on the
fact that it is roughly aligned with the pulsar spin axis and that the
derived energy flux through the jet approximately equals the pulsar spindown
luminosity. They also derive the properties of the medium exterior to the
region in the same way: $kT\sub{ext}=0.12\un{keV}{}$,
$n\sub{ext}=0.16\dhat^{-1/2}\un{cm}{-3}$.

The difficulty with the jet interpretation is that there is a momentum
problem in the jet: the pulsar spindown energy loss rate is entirely
in the form of electromagnetic energy (Pointing flux) and relativistic
particles. Given the energy loss rate, $L\sub{sd}$, this means we know
the momentum loss rate to be simply $P\sub{sd}=L\sub{sd}/c$.
Let us assume that we
somehow solve the problem of converting this roughly spherically symmetric
emission of momentum into a directional flow of momentum that then pushes the
jet. Conservation of momentum then imposes the requirement that
$P\sub{jet}<P\sub{sd}$. But the derived jet properties violate this
limit, as we can see from 
the energy flow and velocity of the jet which Markwardt \& \"Ogelman derived 
from their work surface calculation. They found that $L\sub{jet}\simeq
L\sub{sd}$ and $v\sub{jet}\simeq1000\un{km}{}\un{s}{-1}$. The momentum
flow through the apparently very sub-relativistic jet would then be
$P\sub{jet}=2L\sub{jet}/v\sub{jet}$.  It follows that 
$P\sub{jet}/P\sub{sd} = 2c/v\sub{jet}\gg1$, i.e.\ the momentum available 
in the pulsar emission falls short by a factor 300 of supplying the
required jet momentum. The bubble therefore cannot be a true jet
powered directly by the pulsar.

Another problem with the parameters of the bubble and the main SNR just
outside it derived by Markwardt \& \"Ogelman is that they find the bubble
to be both hotter and denser than its surroundings. This means that its
expansion into its surroundings should be almost like expansion into
vacuum, i.e.\ proceed at the internal sound speed of the bubble, which is
about 500\un{km}{}\un{s}{-1}. But we know from its size and age that the
mean expansion velocity of the bubble has only been about
$50\dhat\un{km}{}\un{s}{-1}$, so unless we live at a very special time
where the bubble has only just become hot, this cannot be. The resolution
for this conundrum was suggested to us by R. McCray, who noted that at
these densities and temperatures the cooling is dominated by a forest of
emission lines around 1\,keV, which create the impression of a thermal peak
at that energy but are not resolvable by ROSAT. The true density can easily
be 10--30 times less, i.e.\ well below that of the main SNR. Hence the
total mass of is of order $0.003\dhat^{5/2}\un{\msun}{}$, and the total
thermal energy content $10^{46}\dhat^{5/2}\un{erg}{}$.  The main SNR
just outside the bubble is too cool for line emission to be significant and
the density is probably close to the value inferred by Markwardt and \"Ogelman.

What we propose is that the X-ray bubble is material from near the core of
the exploding star that was hit by the reverse shock and thus not
accelerated away strongly from the newborn pulsar. The pulsar spindown
energy is absorbed by it close to the pulsar itself, and conducted
throughout the bubble, powering its expansion. We discuss the formation and
evolution of this bubble in \sect{bubble} and propose a binary companion as
the reason for its asymmetric expansion. In \sect{orbit}\ we discuss the
constraints on the progenitor binary that follow from the model.

   \section{The fallback bubble}
   \label{bubble}

The bubble was formed when some gas in between the fallback material and
the fast ejecta stalled (\Fig{fi:cartoon}, panel a). It was then hit by the
energy flux from the newborn pulsar and started pushing into the bulk of
the ejecta (\Fig{fi:cartoon}, panel b). At the same time,  the former
binary companion to the pulsar ploughed through the ejecta, carving a
low-density channel into it. Initially, this channel may cave in behind it
due to the high external pressure. At some point, the pulsar-powered bubble
catches up with the open part of the trail and starts expanding into it
rapidly (panel c). When it catches (almost) up with the companion it has to
adjust its speed to not overtake the companion (panel d). The trail behind
the pulsar now has a somewhat tapered shape, because the expansion of the 
trail due to the bubble pressure has acted longer at the base, making
it wider there. We now discuss the various stages in this scenario in more
detail.
\begin{figure*}
\begin{minipage}[b]{0.75\textwidth}
   {\Large\it see file }{\Large\tt vela\_fg2.gif}

   \vspace*{1cm}
   \end{minipage}\hfill\begin{minipage}[b]{0.23\textwidth}
\caption{Sketches of the companion-punctured bubble scenario. (a) The
         stalled bubble is formed. (b) The bubble is driven into the main
         SNR while the companion starts blazing a trial to the SSW.  (c)
         The bubble catches up with the open part of the companion trail
         and accelerates into it. (d) Present situation, with the bubble
         having caught up with the companion.
         \label{fi:cartoon}
         }
	 \end{minipage}
\end{figure*}

      \subsection{Origin and composition}
      \label{bubborig}

Current models of supernova explosions (Woosley et~al.\ 1994)
\nocite{wwmhm:94} predict that about
$0.03 - 0.1 \msun $ of material near the core boundary will be stalled by
the reverse shock in the envelope and undergo r--processing before escaping
in a neutrino--powered wind. The wind in turn stalls at speeds $\sim
100\un{km}{}\un{s}{-1} $, much less than the bulk of the envelope, if the
initial main-sequence mass of the star is in the narrow range $20-25\msun$
(\Fig{fi:cartoon}a).
The model predicts that the innermost ${\rm few}\times 10^{-5} \msun $ of
the original bubble are composed of about 80\% He and 20\% mixture of
r--process isotopes of atomic number 100--200, and possibly a significant
fraction of isotopes of atomic number near $80$--$90$, like Sr, Y and Zr.
The bubble material should contain about $10^{-5} \msun $ of isotopes
ranging from germanium to lead, including platinum group metals, krypton,
xenon and the rare earths, as wells as molybdenum, tungsten and tin. The
rest of the bubble material comes from near the remnant mass cut, and is
probably chiefly nickel, copper, zinc, calcium, titanium and vanadium as
well as helium.  In this bubble we therefore observe almost directly the
conditions inside an exploding star. Since the outer layers spend
significant time at higher velocities at early times in the supernova,
while the wind--driven r--processed region nearly stalls at late times, we
expect the material from the inner layers not to be well mixed into the
bulk of the supernova remnant.

Since the elemental abundances are sensitive to the explosion models,
spectroscopic detection of any of these isotopes, possibly feasible with
the ASCA satellite or future proposed X--ray satellites, would provide a
test of explosive nucleosynthesis in supernova models. An attempt has been
made to detect $r$-process elements in absorption in the Vela remnant
(Wallerstein et~al.\ 1995)\nocite{wvjf:95} without success.  The lines of
 sight probed were outside the bubble, constrained by the chance
positioning of background B-type stars.  If the hotter bubble  is
over-abundant in $r$-process elements, as conjectured here,
intermediately ionised species might be observed in emission.  Unfortunately
line strengths, or even wavelengths, for highly ionised isotopes beyond
nickel are not available in the literature. One exception is the recently
identified Kr IV line at 534.6 nm (P\'equignot \& Baluteau
1994))\nocite{pb:94}; it is also possible that L or M shell emission
features of high-Z elements could be detected by ASCA.

      \subsection{Evolution}
      \label{bubbevol}

Once the bubble has formed, it will initially cool both due to adiabatic
expansion and radiative cooling, until heating by the pulsar spindown
energy becomes important. There is evidence in the radio structure 1 arcmin
North of the pulsar of a wind termination shock (Bietenholz et~al.\
1991)\nocite{bfh:91} indicating that the pulsar's spindown energy is
absorbed in the bubble close to the pulsar. Ordinary thermal conduction
suffices to transport the pulsar spindown luminosity across the bubble.
The total luminosity of the nebula is less than a percent of $L\sub{sd}$,
so most of the energy goes into heating and expanding the bubble. Indeed, the
cooling time of the bubble is now 100\,Myr, much greater than its age; its
thermal energy ($10^{46}$\,erg) is much less than 
$L\sub{sd}\tau\sub{sd}\simeq2\times10^{48}\un{erg}{}$,
so the bulk of the injected spindown energy is not in the bubble now.
In the past, the cooling time was much shorter than the expansion time
because the cooling time of a constant mass of material scales very steeply
with radius. Then most of the absorbed spindown energy was immediately
radiated, as is the case now in the ten times younger Crab Nebula.

The expansion of the bubble into the surrounding supernova remnant is akin
to the supernova remnant's expansion itself (\Fig{fi:cartoon}b),
 and self-similar solutions for
it exist in various regimes (see review by Ostriker \& McKee 
1986\nocite{om:86}). The problem is more complicated than that of the
supernova explosion itself because the ambient medium is expanding so the
bubble blast wave encounters a decreasing density with time and must
asymptotically match onto a finite expansion velocity.  The energy input is
not all at the beginning as with a standard supernova explosion, but the
pulsar's dipole spindown luminosity decreases with time complicating the
solution further.

The slightly simpler problem of constant ambient density has been solved
and we shall use it to get an idea of the solutions. For a constant rate
of energy injection $L\sub{in}$ the flow enters a self-similar phase once
the initial acceleration phase is over. The shape of $R(t)$ depends on
whether radiative cooling is more important than $p{\rm d}V$ work or not.
Currently, radiative cooling is negligible: the cooling time scale of
100\un{Myr}{} greatly exceeds the expansion time scale, which must be of order
10\,000\un{yr}{}, the bubble's age. Since the (bremsstrahlung) cooling time
of a constant mass of gas scales as $R^{-3}$, and the expansion time scale
is a weak function of $R$, the cross-over from radiative to adiabatic
expansion occurred when the bubble was 10 times smaller than it is now.
Before that time the pulsar energy input was just radiated away and 
little power was applied to driving the expansion. 

The expansion velocity of the bubble near the pulsar can be estimated from
its size and the pulsar age to be about $50\dhat\un{km}{}\un{s}{-1}$.
Since its internal pressure greatly exceeds that of the surrounding ejecta
its expansion is best described by a self-similar power-law similar to the
Sedov-Taylor solution for an adiabatically expanding blast wave, modified
for continuous energy injection (Weaver et~al.\ 1977)\nocite{wmcsm:77}. 

The pressure in the main SNR decreases as $t^{-2}$, faster than the
pressure inside the bubble at early times. Approximately $10^3$ years after
the supernova explosion the pressure inside the bubble exceeds that of the
main SNR and it starts expanding into the main SNR. The initial velocity
profile depends on the exact ratio of the cooling time to the pulsar power
injection at that time (which is uncertain); if the radiative cooling time
is long at that point the bubble material accelerates until it reaches the
self--similar expansion dictated by the balance of work done on the ambient
medium and the instantaneous energy injection; if the initial expansion
velocity is too large, the bubble decelerates to match the self--similar
expansion profile.

The bubble is Rayleigh--Taylor unstable, with mixing time scale much
shorter than the crossing time for the bubble, so material swept up by the
expansion is well mixed through the bubble (Koo \& McKee 1990\nocite{km:90}).
But given the small mass of the bubble, a substantial zone of material
now still outside it should also have come from deep inside the nuclear
cauldron and bear the signs of advanced nucleosynthesis, so the bubble
composition will still be predominantly heavy elements and helium.

      \subsection{Asymmetric expansion}
      \label{bubbasym}

The expansion of the bubble should be roughly spherical if the density of
the surrounding ejecta were uniform, but instead we see a very asymmetric
expansion. Since the blast wave velocity scales as $\rho\sub{ext}^{-3/5}$
(Weaver et~al.\ 1977)\nocite{wmcsm:77}, the density of the ejecta to the
SSW would need to be 20--30 times less than in other directions, which we
feel is rather unlikely to arise by itself. We propose that the asymmetric
expansion follows the path along which the former binary companion of the
pulsar flew away after the supernova explosion disrupted the binary. The
companion moves highly supersonically through the surrounding ejecta and
its wind (either relativistic particles from another pulsar or a normal
stellar wind, see sect.~\ref{orbit}) ploughs a bow shock through the
ejecta, leaving a very low-density region in its wake
(\Fig{fi:cartoon}b).  Such structures of
bow shocks and trails are now well documented near a number of pulsars,
such as the H$\alpha$ nebula of PSR\,1957+20 (Kulkarni \& Hester 1988)
\nocite{kh:88}, the Guitar Nebula (Cordes, Romani, \& Lundgren
1993)\nocite{crl:93} and the X-ray trail of PSR\,1929+10 (Wang, Li, \&
Begelman 1993)\nocite{wlb:93}.  Analogous ideas were proposed for an
apparently fast-moving feature in early observations of SN 1987a (Rees
1987)\nocite{rees:87} and the `chimney' of the Crab nebula (Blandford,
Kennel, \& McKee 1983)\nocite{bkm:83}, and an extensive model along these
lines was proposed for an elongated feature emanating from a supernova
remnant near the Galactic Centre (Nicholls \& Le Strange
1995)\nocite{nl:95}.  The bubble material will expand into this channel
much more rapidly than into the rest of the ejecta (\Fig{fi:cartoon}c).
 The channel will
initially be less than 0.1\un{pc}{}\ in width, but once the bubble
material has filled it, it will expand due to thermal pressure at the same
rate as the rest of the bubble.  This is consistent with the fact that the
extension is as wide as the bubble near its base, and narrows towards the
end.

The width of the path may be estimated by noting that it must be at least
twice the standoff distance of the bow shock that separates the wind of
the moving former binary companion from the exterior material.
(It can be substantially greater, see Nicholls \& Le Strange 1995)\nocite{nl:95}.
For a relativistic wind, as expected from a pulsar, this is
\begin{eqnarray}
  \label{eq:stpuls}
  R\sub{off} = 0.02\un{pc}{}
          &&\left(\frac{\dot{E}}{10^{35}\un{erg}{}\un{s}{-1}}\right)^{1/2}\times
          \nonumber \\
          &&     \left(\frac{n\sub{ext}}{0.16\un{cm}{-3}}\right)^{-1/2}
               \left(\frac{v\sub{c}}{220\un{km}{}\un{s}{-1}}\right)^{-1}.
\end{eqnarray}
($\dot{E}$ is the spindown luminosity, $n\sub{ext}$ the particle number density
outside the bubble, and $v\sub{c}$ the former companion's speed.)
For a stellar wind of velocity $v\sub{w}$ and total mass loss rate
$\dot{M}$, it becomes
\begin{eqnarray}
  \label{eq:ststar}
  R\sub{off} = 0.01\un{pc}{}
          &&\left(\frac{\dot{M}}{10^{-10}\un{\msun}{}\un{yr}{-1}}\right)^{1/2}
               \left(\frac{v\sub{w}}{400\un{km}{}\un{s}{-1}}\right)^{1/2}\times
               \nonumber \\
             &&  \left(\frac{n\sub{ext}}{0.16\un{cm}{-3}}\right)^{-1/2}
               \left(\frac{v\sub{c}}{220\un{km}{}\un{s}{-1}}\right)^{-1}.
\end{eqnarray}
In both cases, we have normalised to the highest likely values of the
energy or mass loss rate (a very vigorous recycled pulsar or extremely
active low-mass star), so one can see that the original channel is unlikely
to be more than about 0.1\un{pc}{}\ wide, rather less than the current
width of 0.25$\dhat$\un{pc}{}\ at the narrowest point. Note that the standoff
distance ahead of the Vela pulsar itself (0.25$\dhat$\un{pc}{}) is in good
agreement with the value computed from \Eq{eq:stpuls}, indicating that the
value used for the external density is reasonable.

The speed of the flow down the channel can be self-regulated to the
companion velocity (\Fig{fi:cartoon}d):
  obviously, if the bubble catches up with the
companion it will run into unshocked ejecta and its expansion will slow
down to that of the rest of the bubble. If it falls behind, the density it
sees ahead of it becomes lower and it is heated more strongly via
conduction from the rest of the bubble, causing its speed to increase
again. As long as heat conduction is sufficient to keep powering the
expansion and the companion speed is lower than the maximum expansion
speed of the bubble (of order its sound speed), it will therefore keep up
with the moving former companion. Both these requirements are fulfilled
now, but as $L\sub{sd}$ decreases and the density and temperature of the
bubble go down, the bubble expansion will start falling behind the
companion motion. We therefore predict that a fast-moving former companion
to the Vela pulsar should be found ahead of the bubble.  Its distance
to the Vela pulsar implies a projected velocity of 
275$\dhat$\un{km}{}\un{s}{-1}\ or a proper motion of 0.25\,arcsec/yr.

   \section{The binary companion and the pre-explosion orbit}
   \label{orbit}

\begin{figure*}
\epsfxsize=0.9\textwidth\epsfbox{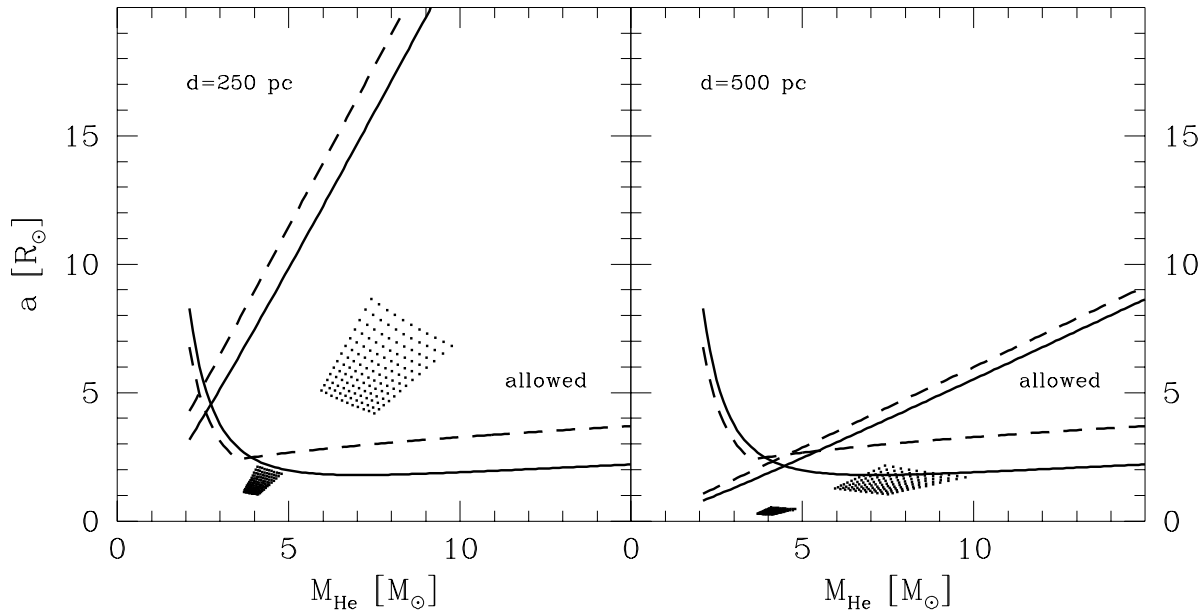}
\caption{Constraints on the progenitor binary. $M\sub{He}$ is the mass of
    progenitor of the Vela pulsar, and $a$ the binary orbital 
    separation. The solid curves apply to a 1.4\msun\ neutron star
    companion, and the dashed ones to an 0.5 \msun\ main-sequence
    companion. Diagonal lines to upper right represent the maximum
    orbital separation consistent with a companion velocity of 
    250$\dhat$\un{km}{}\un{s}{-1}.
    The lower curves represent the minimum separation for which
    neither star in the progenitor binary fills its Roche lobe.
    The left panel is for the nearby distance to Vela of 250\,pc,
    the right one for the classical 500\,pc distance.
    \label{fi:orbit}
    }
\end{figure*}

The pulsar velocity is known from its proper motion to be
$70\dhat\un{km}{}\un{s}{-1}$ (Bailes et~al.\ 1990)\nocite{bmknr:90}
 and in our model the velocity
of its former companion is also known from its position just ahead of the
bubble.  This tightly constrains the initial mass of the Vela pulsar
progenitor and the size of the binary orbit (fig.~\ref{fi:orbit}).  First,
while the pulsar may receive a velocity kick at birth, this is not true for
the companion.  Therefore, energy conservation implies that its orbital
velocity in the progenitor binary must have exceeded
$275\dhat\un{km}{}\un{s}{-1}$. This excludes a red giant as the progenitor
to the supernova, because such a speed cannot be obtained in an orbit wide
enough to accommodate a red giant.  The solid and dashed curves from lower
left to upper right are the orbital separation for which a companion of 1.4
and 0.5\msun, respectively, have an orbital velocity of
$275\dhat\un{km}{}\un{s}{-1}$.  These are firm upper limits to the allowed
orbital separation, both because the companion will use up some of its
velocity in escaping the attraction of the Vela pulsar after the explosion
and because the $275\dhat\un{km}{}\un{s}{-1}$ only represents the companion
speed in the plane of the sky; its true speed must be greater.

The only likely candidate progenitors of the Vela pulsar that fit into the
required orbits of a few solar radii are helium stars or even more evolved
stellar cores. Hence the explosion that formed the Vela supernova remnant
must have been a type Ib or Ic supernova, and any material that can be
identified as purely ejecta should not contain significant amounts of
hydrogen. 
The remnant does contain many filaments that emit H$\alpha$
(Elliot, Goudis, \& Meaburn 1976)\nocite{egm:76}, but that could
well be swept-up material, the total mass of which is by now expected to
exceed the original ejecta mass considerably. It would be quite difficult,
therefore, to determine the supernova type from the ejecta composition 
in a remnant as old as Vela. Especially distinguishing types II and Ib/c,
as we would want, is very difficult in any case because the amounts of 
heavy elements ejected in both would be almost the same.

Close binary stars with pure helium star companions form when a
red giant in a wide binary engulfs its low-mass companion star, which then
plunges into the giant's envelope; this causes the envelope to be ejected
and the orbit to tighten (Bhattacharya \& Van den Heuvel 1991)\nocite{bh:91}.
 The now close binary consists of
the low-mass companion and the core of the red giant, i.e., a naked helium
star.  The two likely companion types are another neutron star, in which
case the pre-explosion binary was probably a strong X-ray source like
Cygnus\,X-3 (Van Kerkwijk et~al.\ 1992, Van den Heuvel \& De Loore
1973)\nocite{kcgkm:92,hd:73}; or a low-mass main-sequence star, in
which case the binary  would have evolved into a low-mass X-ray binary had
it not been disrupted in the supernova explosion.

In both cases, the Vela pulsar binary forms an important link in our
understanding of binary stellar evolution, so we shall investigate them
further.  A lower limit to the size of the orbit follows from the
requirement that neither star overfill its Roche lobe prior to the
explosion (for the radius of the helium star, we took values at core
Carbon ignition from Habets 1986\nocite{habet:86}).
This limit is shown as the lower curve
in fig.~\ref{fi:orbit} for both cases.

If the pulsar received no velocity kick at birth, the mass of the helium
star, $M\sub{He}$, and orbital radius, $a$, follow from the current
velocities of the pulsar and the companion, given their current masses:
 as shown elegantly by Radhakrishnan and Shukre (1985)\nocite{rs:85}, the
velocities at infinity of the binary components after the explosion can be
expressed in terms of the pre-explosion masses $M\sub{He}$ and
$M\sub{comp}$, the mass $\Delta M$ lost in the explosion, and the initial
semi-major axis $a$.  If we now choose $M\sub{comp}$ and fix the pulsar
mass $M\sub{He}-\Delta M$ at 1.4\msun, we are left with two unknowns,
$M\sub{He}$ and $a$, expressed in terms of the measured $v\sub{PSR}$ and
$v\sub{comp}$.  The only uncertainty we have is that the measured
proper-motion velocities depend on the uncertain distance to the Vela
pulsar and that we do not know how much larger the true velocities of the
two objects are than their projections on the plane of the sky; typically,
the ratio of the true to projected velocity would be $\sqrt{3/2}$.  To
account for this, we adopt velocity ranges of
$v\sub{PSR}=$70--100$\dhat$\un{km}{}\un{s}{-1}\ for the pulsar and
$v\sub{comp}=$275--400$\dhat$\un{km}{}\un{s}{-1}\ for the companion.

The smaller group of dots in each panel of Fig.~\ref{fi:orbit} represents
the domain of pre-explosion orbits consistent with those velocity ranges
for the case of a 0.5\msun\ main-sequence companion. 
It lies entirely below the Roche limit (lower dashed curve) for either
distance to Vela and can be ruled out, unless the Vela pulsar 
received a kick velocity at birth of at least 50\un{km}{}\un{s}{-1}.

Possible orbital solutions for the case of a neutron star companion are
indicated by the larger dotted area in \Fig{fi:orbit}. These all fall
above the Roche limit, so a neutron star is a viable companion candidate.
One would expect it to be a recycled pulsar, like the Hulse-Taylor
relativistic binary pulsar, with a magnetic field of perhaps $10^{10}\un{G}{}$
and a period of several tens of milliseconds. It may be a challenge to
detect it as a radio pulsar, given that it is in such a radio bright
region and may even be beamed away from us. But since it was accreting
material until 10,000 years ago, it may still be hot enough be detectable
with some effort in soft X rays or optical radiation, like the Vela pulsar
itself.

Note also that the presence of significant fallback material to form the
bubble constrained the initial mass of the exploding star to be
20--25\msun.  Such a star would leave a naked Helium star of 6--8\msun. 
This overlaps nicely with the range of allowed pre-explosion He star 
masses for the case of a neutron star companion.

   \section{Discussion and Conclusion}

If our model is correct, then the evolution of the Vela remnant and its
surroundings could be rather more complicated than that of a simple
supernova remnant expansion, especially if the companion star is a neutron
star, for then a prior supernova exploded at the location of the Vela
remnant 3--10\,Myr ago, and the Vela SNR is now expanding into this old
remnant.  In fact, there is a region of about 20\degs\ radius around the
Vela SNR called the Gum Nebula, which is tenuous and ionised and could well
be either an old wind bubble or SNR.  Somewhere along its way, the Vela SNR
would overtake the envelope of the red giant that was ejected during the
spiral-in of the first-born neutron star.  Since the Gum Nebula as a whole
is rather tenuous, this might well change the expansion rate significantly.
The speed of the ejected envelope is probably about 100\un{km}{}\un{s}{-1},
and the He star would live for another few hundred thousand years after the
spiral-in, so it would be a few tens of pc away when the Vela pulsar was
formed. Since the radius of the Vela SNR is 15$\dhat$\,pc, the shell could
be ploughing into this material now.  This could seriously confuse
estimates of the age of the remnant based on the current size and X-ray
brightness (Aschenbach et~al.\ 1995)\nocite{aet:95}.

We have shown that the newly discovered asymmetric X-ray feature near the
Vela pulsar can be interpreted as a pulsar-powered bubble of fallback
material that originated very near the collapsing core in the supernova
explosion that formed the Vela pulsar. The asymmetry of the expansion can
be explained by the hypothesis that the Vela pulsar had a close binary
companion before the explosion which is now ploughing a channel though the
Vela remnant into which the bubble can expand much more rapidly than in
other directions.

The derived orbital radius and mass of the progenitor binary imply that
Vela is the remnant of a type Ib supernova, and that either the companion
is a neutron star or that the Vela pulsar received a velocity kick at
birth.  Our model is easily falsifiable because it makes a number of very
specific predictions: the bubble should contain large amounts of heavy
r-process elements, and at or near the end of the elongated X-ray bubble
there should be a former binary companion to the Vela pulsar, most likely
a neutron star or a late-type main-sequence star. It can easily be
recognised as the former binary companion because of the direction and
magnitude of its proper motion ($0\farcs25$/yr towards the
south-southwest). If such observations confirm our model, the chemical
composition of the bubble will provide an rare and highly valuable
test of explosive nucleosynthesis in supernovae.

 \section*{Acknowledgements}

We are grateful to Niel Brandt, Melvyn Davies, Dick McCray, Jim Pringle,
Martin Rees, Mal Ruderman, Chris Thompson, and Jacco Vink for helpful
discussions.  We thank C. Markwardt and B. Aschenbach for providing us
with the basic figures from which figure 1 was created, and R. Sword for
creating figures 1 and 2.  This work was supported by PPARC fellowships to
RAMJW and SS.


\end{document}